\documentclass[conference]{IEEEtran}

\IEEEoverridecommandlockouts

\usepackage{cite}
\usepackage{amsmath,amssymb,amsfonts}
\usepackage{algorithmic}
\usepackage{graphicx}
\usepackage{textcomp}
\def\BibTeX{{\rm B\kern-.05em{\sc i\kern-.025em b}\kern-.08em
    T\kern-.1667em\lower.7ex\hbox{E}\kern-.125emX}}

\usepackage{tikz}
\usetikzlibrary{spy}
\usetikzlibrary{calc,arrows.meta}
\usepackage{pgfplots}
\usetikzlibrary{plotmarks}
\usetikzlibrary{arrows.meta}
\usetikzlibrary{positioning,fit}
\usetikzlibrary{intersections}
\usetikzlibrary{patterns}

\usepgfplotslibrary{patchplots}

\pgfdeclarelayer{back1}
\pgfdeclarelayer{lay1}
\pgfdeclarelayer{back2}
\pgfdeclarelayer{lay2}
\pgfsetlayers{back2,lay2,back1,lay1,main}

\newcommand{\exportFigures}{false}
\newcommand{\exportFiguresAsPNG}{true}

\ifthenelse{\equal{\exportFigures}{true}}
{
  \usepgfplotslibrary{external}
  \tikzexternalize[prefix=compiled_tikz_figures/]
  \ifthenelse{\equal{\exportFiguresAsPNG}{true}}
  {
    \tikzset
    {   png export/.style={
        external/system call={
        pdflatex \tikzexternalcheckshellescape -halt-on-error --extra-mem-top=10000000 -interaction=batchmode -jobname "\image" "\texsource" && pdftops -eps "\image.pdf" && convert -density 700 -transparent white "\image.pdf" "\image.png"
    }}}
  \tikzset{png export}
  }
  {}
}
{}

\usepackage{url}

\usepackage{bm}
\usepackage[caption=false,font=footnotesize]{subfig}
\usepackage[normalem]{ulem}

\usepackage{mathtools, cuted}

\usepackage{acronym}

\usepackage{booktabs}

\usepackage[latin1]{inputenc}
\usepackage{tikz}
\usetikzlibrary{shapes,arrows}
\usetikzlibrary{arrows.meta}
\usetikzlibrary{positioning}

\usepackage{ellipsis}
\usetikzlibrary{calc}
\usetikzlibrary{decorations.pathreplacing,decorations.markings,shapes.geometric}
\usetikzlibrary{decorations.pathmorphing}
\usetikzlibrary{fit}

\usepackage[most]{tcolorbox}
\tcbuselibrary{breakable}
\tcbset{every box/.style={enhanced,breakable}}
\tcbset{colframe=black,colback=red!10,enhanced,breakable,sharp corners}

\usepackage{enumitem} 

\usepackage{pdfpages}

\definecolor{alex}{RGB}{51,183,150}
\definecolor{erik}{RGB}{235,134,52}

\newcommand{\ticked}{$\text{\rlap{$\checkmark$}}\square$}
\newcommand{\unticked}{{$\square$}}
\newcommand{\tick}[1]{\ifthenelse{#1=1}{\ticked}{\unticked}}

\newcommand{\rmv}{\hspace*{-.3mm}}

\newcommand{\E}[1]{\ensuremath{\mathbb{E}\!\left[#1\right]}}

\newcommand{\norm}[2]{\ensuremath{\lVert #1 \rVert^{#2}}}

\newcommand{\minus}{\rmv - \rmv}

\newcommand{\s}{\hspace*{0.5pt}}

\newcommand{\pe}{p_{\text{E}\s n}^{(j)}(u_n^{(j)}\rmv\rmv\rmv,q_n^{(j)})}
\newcommand{\pd}{p_{\text{D}\s n}^{(j)}(u_n^{(j)})}
\newcommand{\sigmadhat}{\hat{\sigma}_{\text{d}\s n,m}^{(j)}}
\newcommand{\dlos}{d^{(j)}_{\text{LOS}\s n}(\bm{p}_n)}
\newcommand{\sigmaahat}{\hat{\sigma}_{\mathrm{\alpha}\s n,m}^{(j)}}

\newlength{\figureheight}
\newlength{\figurewidth}

\allowdisplaybreaks

\definecolor{mycolor01}{rgb}{0.00000,0.00000,1.00000}
\definecolor{mycolor02}{rgb}{0.133,0.545,0.133}
\definecolor{mycolor03}{rgb}{0.50000,0.00000,0.50000}
\definecolor{mycolor04}{rgb}{1.00000,0.83984,0.00000}
\definecolor{mycolor05}{rgb}{0.92969,0.50781,0.92969}
\definecolor{mycolor06}{rgb}{1.00000,0.64453,0.00000}
\definecolor{mycolor07}{rgb}{0.50000,0.50000,0.50000}
\definecolor{mycolor08}{rgb}{1.00000,0.00000,0.00000}
\definecolor{mycolor09}{rgb}{0.00000,0.50000,0.00000}
\definecolor{mycolor10}{rgb}{0.54297,0.00000,0.00000}

\tikzset{
  nomorepostactions/.code={\let\tikz@postactions=\pgfutil@empty},
  decmark/.style 2 args={decoration={markings,
    mark= between positions 0 and 1 step (1/6)*\pgfdecoratedpathlength with{%
        \tikzset{#2,every mark}\tikz@options
        \pgftransformresetnontranslations
        \pgfuseplotmark{#1}%
      },  
    },
    postaction={decorate},
    /pgfplots/legend image post style={
        mark=#1,mark options={#2},every path/.append style={nomorepostactions}
    },
  },
}

\pgfplotsset{
resultStyle1/.style={mark=none, line width=0.5pt, mycolor01, decmark={oplus}{solid}},
resultStyle2/.style={mark=none, line width=0.5pt, mycolor02, decmark={+}{solid}},
resultStyle3/.style={mark=none ,line width=0.5pt, mycolor03, decmark={triangle}{solid}},
resultStyle4/.style={mark=none, line width=0.5pt, mycolor06, decmark={star}{solid}},
resultStyle5/.style={mark=none, line width=0.5pt, mycolor08, decmark={o}{solid}},
resultStyle6/.style={mark=none, line width=0.5pt, mycolor09, decmark={square}{solid}}, 
resultStyleBase/.style={mark=none, line width=0.5pt,}, 
}
  
  \pgfplotsset{
        compat=newest,
        simple style/.style={
                label style={font=\scriptsize},
                legend style={font=\scriptsize},
                tick label style={font=\scriptsize},
                nodes near coords style={font=\scriptsize},
                title style={font=\scriptsize},
                width=\figurewidth,
                height=\figureheight,
                at={(0\figurewidth,0\figureheight)},
                scale only axis,
                grid style={dotted},
                mark options={solid}, 
        },
        base style/.style={
                label style={font=\scriptsize},
                legend style={font=\scriptsize},
                tick label style={font=\scriptsize},
                nodes near coords style={font=\scriptsize},
                title style={font=\scriptsize},
                width=\figurewidth,
                height=\figureheight,
                at={(0\figurewidth,0\figureheight)},
                scale only axis,
                cycle list={
                {mark=none, line width=0.5pt, mycolor01, solid},
                {mark=none, line width=0.5pt, mycolor02, dash dot},
                {mark=none ,line width=0.5pt, mycolor03, densely dashed},
                {mark=none, line width=0.5pt, mycolor04, dash dot dot},
                {mark=x   , line width=0.5pt, mycolor05},
                {mark=.   , line width=0.7pt, mycolor06}, 
                {mark=square,only marks, mark size = 0.8pt, mycolor07,
                mark options = {line width = 0.4pt}},
                {mark=x,     only marks, mark size = 1.3pt, mycolor08,
                mark options = {line width = 0.4pt}},
                {mark=o,     only marks, mark size = 0.8pt, mycolor09,
                mark options = {line width = 0.4pt}},
                {mark=o, mycolor10},
                },
                grid style={dotted},
                xmajorgrids,
                ymajorgrids,
                mark options={solid}, 
        },
        std graph style new/.style={
                xlabel style={yshift=1mm},
                ylabel style={yshift=-1.5mm},
                yticklabel style={xshift=1mm},
        },
        pdf graph style/.style={
                xlabel style={yshift=1mm},
                ylabel style={yshift=-1.5mm},
                yticklabel style={xshift=1mm},
                xmajorgrids,
                ymajorgrids,
                mark repeat = 1,
                mark phase = 0,
                cycle list={
                    {mark=none, solid,line width=0.5pt, mycolor01},
                    {mark=none, line width=0.7pt, mycolor02}, 
                    {mark=none, dash dot,line width=0.5pt, mycolor06},
                    {mark=none, densely dashed,line width=0.5pt, mycolor05},
                    {mark=none, dash dot dot,line width=0.5pt, mycolor03},
                    {mark=none, line width=0.7pt, mycolor06}, 
                    {mark=square,only marks, mark size = 0.6pt, mycolor07},
                    {mark=x,     only marks, mark size = 0.9pt, mycolor08},
                    {mark=*,     only marks, mark size = 0.6pt, mycolor09},
                    {mark=none, black, dashed, forget plot},
                    {mark=none, black, dashed, forget plot},
                    {mark=none, black, dashed, forget plot},
                },
                ytick = {0, 0.5e-2, 1e-2, 1.5e-2},
                yticklabels = {$0$, $0.5$, $1$, $1.5$},
                ylabel={PDF},
        },
        meas graph style/.style={
                xlabel style={yshift=1mm},
                ylabel style={yshift=-1mm},
                xmajorgrids,
                ymajorgrids,
                mark repeat = 1,
                mark phase = 0,
                cycle list={
                    {color=black, only marks, mark=*, mark size=0.5pt, mark options={solid, black}},
                    {color=red, only marks, mark=*, mark size=0.1pt, line width=0.25pt},
                },
                ylabel={},
        }, 
        ci graph style/.style={
                xlabel style={yshift=1mm},
                ylabel style={yshift=-1.5mm},
                yticklabel style={xshift=1mm},
                mark repeat = 1,
                mark phase = 0,
                ymin=1e-3,
                ymax=100,
                ytick = {100, 50, 10, 1, 0.1, 0.01, 1e-3, 1e-4},
                yticklabels = {$0$, $50$, $90$, $99$, $99.9$, $99.99$, $99.999$, $99.9999$},
                y dir=reverse,
        },     
        bp coeff style/.style={
               scale only axis=true,
               width=0.225*.9\linewidth,
               height=0.225*.9\linewidth,
               scale only axis,
               xmin=-4.000,
               xmax=4.000,
               xlabel={$\ell${\color{white}$\aod$}},
               ticklabel style={font=\footnotesize},
               ymin=0.000, ymax=0.9,
               ylabel={$c_\ell$},
               xlabel style={font=\footnotesize},
               ylabel style={font=\footnotesize},
               major tick length=2pt
        },
        bp graph style/.style={        
               scale only axis=true,
               width=0.35*1.1\linewidth,
               height=0.225*.9\linewidth,
               scale only axis,
               xmin=-3.14, xmax=3.14,
               xlabel={$\aod${\color{white}$\ell$}},
               ticklabel style={font=\footnotesize},
               xtick={-3.14,-1.57,0.0,1.57,3.14},
               xticklabels={$-\pi$,$-\tfrac{\pi}{2}$,$0$,$\tfrac{\pi}{2}$,$\pi$},
               ymin=0.000, ymax=3,
               ylabel={Beampattern},
               xlabel style={font=\footnotesize}, ylabel style={font=\footnotesize},
               major tick length=2pt
        },
        peb graph style/.style={        
               width=0.66\linewidth,
               scale only axis,
               point meta min=-2.583,
               point meta max=-0.300,
               axis on top,
               xmin=0.000,
               xmax=12.000,
               xlabel={x in meter},
               y dir=reverse,
               ymin=0.000,
               ymax=8.000,
               ylabel={y in meter},
               ytick={7.0,6.0,...,0.0},
               xtick={0.0,1.0,...,12.0},
               yticklabels={$1$,$2$,$3$,$4$,$5$,$6$,$7$,$8$},
               xlabel style={font=\scriptsize,yshift=0.125cm},
               ylabel style={font=\scriptsize,yshift=-0.125cm},
               ticklabel style={font=\scriptsize},
               unit vector ratio*=1 1 1,
               yticklabel pos=left,
               major tick length=2pt,
               colormap={mymap}{[1pt] rgb(0pt)=(1,1,1); rgb(1pt)=(0.858903,0.984776,0.839302); rgb(2pt)=(0.777958,0.94143,0.649487); rgb(3pt)=(0.755504,0.864264,0.463393); rgb(4pt)=(0.777509,0.754439,0.310168); rgb(5pt)=(0.820314,0.619497,0.21003); rgb(6pt)=(0.854796,0.471879,0.170327); rgb(7pt)=(0.851327,0.326629,0.183322); rgb(8pt)=(0.784671,0.198575,0.225774); rgb(9pt)=(0.637629,0.0993149,0.259577); rgb(10pt)=(0.400067,0.0343393,0.229819); rgb(11pt)=(0,0,0)},
               colorbar style={ylabel={Position Error Bound in centimeter (logscale)}, ytick={-0.4,-0.82,...,-2.92}, yticklabels={$39.8$, $15.1$, $5.8$, $2.2$, $0.8$, $0.3$},ylabel style={yshift=0.5mm,font=\scriptsize,scale=0.8},width=2.0mm,xshift=-4.25mm,ticklabel style={font=\scriptsize},major tick length=0pt}, 
               colormap access=piecewise constant
        },
        peb ellipses/.style={color=white, line width=0.4pt, forget plot}
    }

\tikzset{naming/.style={align=center,font=\small}}
\tikzset{antenna/.style={insert path={-- coordinate (ant#1) ++(0,0.25) -- +(135:0.25) + (0,0) -- +(45:0.25)}}}
\tikzset{station/.style={naming,draw,shape=dart,shape border rotate=90, minimum width=10mm, minimum height=10mm,outer sep=0pt,inner sep=3pt}}
\tikzset{mobile/.style={naming,draw,shape=rectangle,minimum width=12mm,minimum height=6mm, outer sep=0pt,inner sep=3pt}}
\tikzset{radiation/.style={{decorate,decoration={expanding waves,angle=90,segment length=4pt}}}}

\tikzset{
  pobl/.style={
    inner sep=0pt, outer sep=0pt, fill=#1,
  },
  pobl gron/.style n args={2}{
    pobl=#1, rounded corners=#2,
  },
  pics/person/.style n args={3}{
    code={
      \node (-corff) [pobl=#1, minimum width=.25*#2, minimum height=.375*#2, rotate=#3, pic actions] {};
      \node (-pen) [minimum width=.3*#2, circle, pobl=#1, outer sep=.01*#2, anchor=south, rotate=#3, pic actions] at (-corff.north) {};
      \node (-coes dde) [pobl gron={#1}{1pt}, anchor=north west, minimum width=.12125*#2, minimum height=.25*#2, rotate=#3, pic actions] at (-corff.south west) {};
      \node [pobl=#1, anchor=north, minimum width=.12125*#2, minimum height=.15*#2, rotate=#3, pic actions] at (-coes dde.north) {};
      \node (-coes chwith) [pobl gron={#1}{1pt}, anchor=north east, minimum width=.12125*#2, minimum height=.25*#2, rotate=#3, pic actions] at (-corff.south east) {};
      \node [pobl=#1, anchor=north, minimum width=.12125*#2, minimum height=.15*#2, rotate=#3, pic actions] at (-coes chwith.north) {};
      \node (-braich dde) [pobl gron={#1}{.75pt}, minimum width=.075*#2, minimum height=.325*#2, outer sep=.0064*#2, anchor=north west, rotate=#3, pic actions] at (-corff.north east)  {};
      \node [pobl=#1, minimum width=.05*#2, minimum height=.2*#2, outer sep=.0064*#2, anchor=north west, rotate=#3, pic actions] at (-corff.north east) {};
      \node (-braich chwith) [pobl gron={#1}{.75pt}, minimum width=.075*#2, minimum height=.325*#2, outer sep=.0064*#2, anchor=north east, rotate=#3, pic actions] at (-corff.north west) {};
      \node [pobl=#1, minimum width=.0375*#2, minimum height=.2*#2, outer sep=.0064*#2, anchor=north east, rotate=#3, pic actions] at (-corff.north west) {};
      \node (-fit person) [fit={(-pen.north) (-braich dde.east) (-coes chwith.south) (-braich chwith.west)}] {};
    },
  },
  pics/SBS/.style={code={
      \begin{scope}[local bounding box=#1]
      \fill [pic actions/.try] (-1,0) -- (-1/2,3) -- (1/2, 3) -- (1,0) -- cycle;
      \fill [pic actions/.try] (-1/16,2) rectangle (1/16,4);
      \fill [pic actions/.try] (0,4) circle [radius=1/4];
      \foreach \i in {-1,1}
        \fill [shift=(90:4), xscale=\i]
          \foreach \r in {1,3/2,2}{
            (-45:\r) arc (-45:45:\r) -- (45:\r-1/10)
            arc(45:-45:\r-1/10) -- cycle
          };
       \end{scope}
  }},
}

\begin{document}


\title{A Message Passing based Adaptive PDA Algorithm for Robust Radio-based Localization and Tracking}


\author{Alexander Venus$^{1,2}$, Erik Leitinger$^{1}$, Stefan Tertinek$^{3}$, and  Klaus Witrisal$^{1,2}$
\thanks{The financial support by the Christian Doppler Research Association, the Austrian Federal Ministry for Digital and Economic Affairs and the National Foundation for Research, Technology and Development is gratefully acknowledged.}


\\
\small{{$^1$Graz University of Technology, Austria}, {$^3$NXP Semiconductors, Austria},}\\
\small{{$^2$Christian Doppler Laboratory for Location-aware Electronic Systems}}\\
}


\maketitle
\frenchspacing
\begin{abstract}

We present a message passing algorithm for localization and tracking in multipath-prone environments that implicitly considers obstructed line-of-sight situations. The proposed adaptive \acl{pda} algorithm infers the position of a mobile agent using multiple anchors by utilizing delay and amplitude of the \acp{mpc} as well as their respective uncertainties. By employing a non-uniform clutter model, we enable the algorithm to facilitate the position information contained in the \acp{mpc} to support the estimation of the agent position without exact knowledge about the environment geometry. Our algorithm adapts in an online manner to both, the time-varying \acl{snr} and \ac{los} existence probability of each anchor.
In a numerical analysis we show that the algorithm is able to operate reliably in environments characterized by strong multipath propagation, even if a temporary obstruction of all anchors occurs simultaneously.
%
\end{abstract}


\IEEEpeerreviewmaketitle

\acresetall



\section{Introduction}\label{sec:introduction}

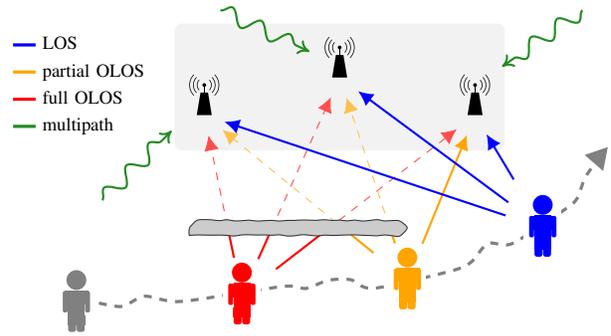
\begin{figure}[t]
 \vspace{-1mm}
 \centering
 \newcounter{randymark}
\pgfdeclaredecoration{mark random y steps}{start}
{%
  \state{start}[width=+0pt,next state=step,%
  persistent precomputation={\pgfdecoratepathhascornerstrue%
  \setcounter{randymark}{0}}]{
  \stepcounter{randymark}
  \pgfcoordinate{randymark\arabic{randymark}}{\pgfpoint{0pt}{0pt}}
  }%
  \state{step}[auto end on length=1.5\pgfdecorationsegmentlength,
               auto corner on length=1.5\pgfdecorationsegmentlength,
               width=+\pgfdecorationsegmentlength]
  { \stepcounter{randymark}
    \pgfcoordinate{randymark\arabic{randymark}}{\pgfpoint{\pgfdecorationsegmentlength}{rand*\pgfdecorationsegmentamplitude}}
  }%
  \state{final}
  {
    \stepcounter{randymark}
    \pgfcoordinate{randymark\arabic{randymark}}{\pgfpointdecoratedpathlast}}%
}%

\def\sinline{\tikz\draw[decorate,thick,decoration={snake,amplitude=1.0mm, segment length=12pt}] (0,0) -- (1.2,0);}

\hspace*{-0.7cm}
\begin{tikzpicture}[
myarrow/.style= {arrows={-Latex[length=2mm,width=2mm]}, thin, shorten <=2.5mm,shorten >=1.5mm},
myarrowint/.style= {thick, shorten <=2.5mm},
myarrowmp/.style= {shorten >=-3pt,thick, arrows={-{>[width=2mm,length=1.5mm]}}, mycolor02},
patharrow/.style= {black!50, very thick,arrows={-triangle 60[length=2.8mm,width=3mm]}},
box/.style={rounded corners=1mm,draw=black!0!white, fill = black!5, inner xsep = 4mm, inner ysep = 6mm},
leg/.style={font=\scriptsize,rectangle,rounded corners=1mm,draw=black,fill=white,minimum width=2cm,minimum height=2cm,inner sep=2cm},
path/.style={black!50, dashed, very thick},
personbox/.style = {rectangle,fill=white, inner sep=0.25cm},
personbox2/.style = {rectangle,fill=white, inner sep=0.25cm, minimum width=0.5cm,minimum height=1cm},
]



\node (antenna2) at (0+0, 1.0) {};
\node (antenna3) at (1.8+0, 0.5) {};
\node (antenna1) at (-1.8+0, 0.5) {};
\pic [fill=black, scale = 0.10] at (antenna1) {SBS};
\pic [fill=black, scale = 0.10] at (antenna2) {SBS};
\pic [fill=black, scale = 0.10] at (antenna3) {SBS};

\node[personbox] (person1) at (-3.5,-2) {};
\node[personbox] (person2) at (-1.3,-1.9) {};
\node[personbox] (person3) at (1.3-0.4,-1.7) {};
\node[personbox2] (person4) at (2.7,-1) {};
\node (edge) at  (3.5, 0) {};
 
 \begin{pgfonlayer}{back1}
 \path[decoration={mark random y steps,segment length=3mm,amplitude=1mm},decorate] ($(person1)+(0:3mm)$) -- (person2) -- (person3) -- ($(person4)+(0,-0.05)$) -- (edge);
\draw[path] plot[variable=\x,samples at={1,...,\arabic{randymark}},smooth] 
(randymark\x);
 \end{pgfonlayer};
 
%
%
\draw[patharrow]  ($(edge)+(49:0.1mm)$) -- ($(edge)+(49:1mm)$);

\draw pic at (person1) {person={black!50}{25pt}{0}};
\draw pic at (person2) {person={mycolor08}{25pt}{0}};
\draw pic at (person3){person={mycolor06}{25pt}{0}};      
\draw pic at (person4){person={mycolor01}{25pt}{0}};


\begin{pgfonlayer}{back2}
\begin{axis}[hide axis,
    at={(person1)}, 
    width = 7cm,
    height =5cm,
    xmin = -2, xmax=8,
    ymin = -3, ymax=6,
    ticks=none,
    legend style={
    at={(-0.18,1.08)}, anchor=north west, legend cell align=left, align=left, draw=white,
    /tikz/every even column/.append style={column sep= 2mm},
    legend image post style={xscale=0.5, yscale=0.5},
    },
    font={\scriptsize},
    ]
 
\addlegendimage{color=mycolor01,very thick} \label{fig:eye_catcher:los}
\addlegendentry{LOS};
\addlegendimage{color=mycolor06,very thick} \label{fig:eye_catcher:partial_nlos}
\addlegendentry{partial OLOS};
\addlegendimage{color=mycolor08,very thick} \label{fig:eye_catcher:full_nlos}
\addlegendentry{full OLOS};
\addlegendimage{color=mycolor02,very thick} \label{fig:eye_catcher:mp}
\addlegendentry{multipath};
\end{axis}
\end{pgfonlayer};
 
\draw [name path=line r1, myarrow, mycolor08!70, dashed] (person2) -- (antenna1);
\draw [name path=line r2, myarrow, mycolor08!70, dashed] (person2) -- (antenna2);
\draw [name path=line r3, myarrow, mycolor08!70, dashed] (person2) -- (antenna3);

\draw [name path=line o1, myarrow, mycolor06!70, dashed] (person3) -- (antenna1);
\draw [name path=line o2, myarrow, mycolor06!70, dashed] (person3) -- (antenna2);
\draw [myarrow, mycolor06!100, thick] (person3) -- (antenna3);

\draw [myarrow, mycolor01!100, thick] (person4) -- (antenna1);
\draw [myarrow, mycolor01!100, thick] (person4) -- (antenna2);
\draw [myarrow, mycolor01!100, thick] (person4) -- (antenna3);

\node (ob1) at (0,-1){};
\path[name path=line ob]  ($(ob1)-(20mm,1mm)$) -- ($(ob1)+(9mm,1mm)$);

\draw[mycolor08!100, myarrowint, name intersections={of=line r1 and line ob}] (person2) -- (intersection-1);
\draw[mycolor08!100, myarrowint, name intersections={of=line r2 and line ob}] (person2) -- (intersection-1);
\draw[mycolor08!100, myarrowint, name intersections={of=line r3 and line ob}] (person2) -- (intersection-1);

\draw[mycolor06!100, myarrowint, name intersections={of=line o1 and line ob}] (person3) -- (intersection-1);
\draw[mycolor06!100, myarrowint, name intersections={of=line o2 and line ob}] (person3) -- (intersection-1);

\path[decoration={mark random y steps,segment length=1mm,amplitude=0.1mm},decorate] ($(ob1)-(20mm,1mm)$) rectangle ($(ob1)+(9mm,1mm)$);
\draw[draw=black!80!white,fill=black!20!white] plot[variable=\x,samples at={1,...,\arabic{randymark}},smooth](randymark\x);


\begin{pgfonlayer}{back1}
\node[box, fit = (antenna2.north) (antenna1.north) (antenna3.north)] (carbox) {};
\end{pgfonlayer};

\node[myarrowmp,rotate=45] (mp) at ($(antenna1.north)-(0.9,0.8)$){\sinline};
\node[myarrowmp,rotate=-25] (mp2) at ($(antenna2.north)+(-1,0.5)$){\sinline};
\node[myarrowmp,rotate=-145] (mp2) at ($(antenna3.north)+(0.9,0.9)$){\sinline};


\end{tikzpicture}
 \setlength{\abovecaptionskip}{0pt}
 \setlength{\belowcaptionskip}{0pt}
 \caption{The mobile agent \protect\scalebox{0.6}{{\protect\tikz[inner sep=0pt] \protect\draw pic at (0,0) {person={black!50}{15pt}{0}}; }} is walking alongside the anchors \protect\scalebox{0.5}{{\protect\tikz[inner sep=0pt] \protect\pic [fill=black, scale = 0.10] at (0,0) {SBS}; }} on an example trajectory. Due to an obstacle, the LOS to all anchors \protect\ref{fig:eye_catcher:los} is not always available. There occur partial \protect\ref{fig:eye_catcher:partial_nlos}, as well as full OLOS \protect\ref{fig:eye_catcher:full_nlos} situations.
 }\label{fig:eye_catcher}
\end{figure}

Radio-based localization in environments such as indoor or urban territories is still a challenging task\cite{WitrisalSPM2016Copy,ShahmansooriTWC2018Copy}. These environments are characterized by strong multipath propagation and frequent obstructed line-of-sight (OLOS)\acused{olos} situations, which can prevent the correct extraction of the \ac{los} component (see Fig.~\ref{fig:eye_catcher}). 
For safety and security critical applications, such as keyless entry systems \cite{Kalyanaraman2020} or autonomous driving \cite{Karlsson2017}, robustness, i.e., a low probability of localization outage, is of critical importance. 
%

Therefore, new systems take advantage of multipath channels by estimating \acp{mpc} for localization\cite{LeitingerTWC2019,GentnerTWC2016Copy}, exploiting cooperation among agents\cite{WymeerschProc2009}, or signal processing against multipath propagation and clutter measurements, i.e., outliers, in general\cite{WymeerschIEEE2012, LeitingerGNSS2016, MeyerFUSION2018}.

The \acl{pdaf}\cite{BarShalom1995} is a Gaussian variant of \acf{pda}, which is able to incorporate multiple anchors (sensors)\cite{BarShalom1995} and amplitude information (\acs{pdaai}\cite{LerroACC1990}), but suffers from being computationally intractable or its dependence on mode-matching. 


%


This paper proposes a low-complexity message passing based multi-sensor \ac{pda} algorithm that estimates and tracks the state of a mobile agent by utilizing delay and amplitude of \acfp{mpc} as well as their respective uncertainties. 
The proposed algorithm adapts in an online manner to both, the time-varying \ac{snr}\cite{LeitingerTWC2019} and \ac{los} existence probability of each anchor
\cite{SoldiTSP2019}.
Furthermore, we use a non-uniform \ac{nlos} model, which comprises measurements originating from \acp{mpc}, as well as \acp{fa}, which do not have a physical explanation (similar to \cite{Yu2020}).
This model enables the algorithm to utilize the position information contained in the \acp{mpc} in order to support the estimation of the agent state without specific map information and, hence, to operate reliably in environments with strong multipath propagation and temporary obstructed LOS situations. 
In that sense, the proposed algorithm allows to indirectly exploit MPCs.
%
The key contributions of this paper are as follows.
\setlist[itemize]{leftmargin=6mm}
\begin{itemize}
 \item We present a multi-sensor message passing algorithm with combined \ac{snr} and \ac{los} existence probability tracking.
 \item We employ a non-uniform \ac{nlos} \ac{pdf} using a double-exponential model for the multipath \ac{lhf}. 
 \item We show the applicability of our algorithm in the context of \ac{olos} mitigation and evaluate the influence of the features of our algorithm in a numerical analysis. 
\end{itemize}
Note that for this work it is assumed that the parameters of the \ac{nlos} object are known constants. This shortcoming shall be addressed in an extended version of this work, and is further discussed in Section~\ref{sec:conclusion}.


\section{Signal Model}\label{sec:signal_model}


At each discrete time $n \in \{1,\,...\,,N\}$, the mobile agent at position $\bm{p}_n$ transmits a  signal $s(t)$ and each anchor $j \in \{1,\,...\,,J\}$ at anchor position $\bm{p}_{\text{A}}^{(j)} = [p_{\text{Ax}}^{(j)}\; p_{\text{Ay}}^{(j)}]^\text{T}$ acts as a receiver. 
The complex baseband signal received at the $j$th anchor is modeled as
\vspace{-2mm}
\begin{align}\label{eq:signal_model} \vspace{-10mm}
 \rmv\rmv\rmv\rmv\rmv\rmv r_n^{(j)}(t) =  \alpha_{n,0}^{(j)} s\big(t\minus\tau_{n,0}^{(j)}\big) \rmv  + \rmv\rmv\rmv \sum_{k = 1}^{{K}_n^{(j)}}\alpha_{n,k}^{(j)} s\big(t\minus\tau_{n,k}^{(j)}\big) 
 \rmv +  w_{n}^{(j)}(t)
\end{align}
The first and second term describe the \ac{los} component and the sum of ${K}_n^{(j)}$ specular \acp{mpc} with their corresponding complex amplitudes $\alpha_{n,k}^{(j)}$ and delays $\tau_{n,k}^{(j)}$, respectively. The third term $w_{n}^{(j)}(t)$ is \acl{awgn} with double-sided power spectral density $N_0/2$. The \acp{mpc} arise from reflection or scattering by unkown objects, since we assume that no map information is available as indicated in Fig.~\ref{fig:eye_catcher} by green lines.
%

\section{Channel Estimation} \label{sec:channel_estimation}

The received signal \eqref{eq:signal_model} is sampled and, by applying a suitable snapshot based channel estimation and detection algorithm 
\cite{RichterPhD2005,BadiuTSP2017}, one obtains at each time $n$ and anchor $j$, a number of $M_n^{(j)}$ measurements denoted by ${\bm{z}^{(j)}_{n,m}}$, with measurement indices $m \in \mathcal{M}_n^{(j)} = \{1,\,...\,,M_n^{(j)}\}$. Each ${\bm{z}^{(j)}_{n,m}} = [\hat{d}_{n,m}^{(j)}~ \hat{u}^{(j)}_{n,m}~ \sigmadhat]^\text{T}$ contains a distance measurement $\hat{d}_{n,m}^{(j)} = c\, \hat{\tau}_{n,m}^{(j)}$, a normalized signal amplitude measurement $\hat{u}^{(j)}_{n,m}~=~|\hat{\alpha}^{(j)}_{n,m}|/\sigmaahat$ corresponding to the square root of the \ac{snr}, and a distance standard deviation measurement $\sigmadhat$, where $\hat{\tau}_{n,m}^{(j)}$ and $\hat{\alpha}^{(j)}_{n,m}$ represent the corresponding delay and complex amplitude and $c$ is the speed-of-light. 
If not implicitly provided by the channel estimator, we can obtain $\sigmadhat$ by means $\hat{u}^{(j)}_{n,m}$ using
$\sigmadhat =  ( c \,\sqrt{8}\,  \pi \, \beta  \,\hat{u}^{(j)}_{n,m} )^{-1}$, with $\beta$ being the effective bandwidth.
%
This equation corresponds to the \ac{crlb} for a single-distance measurement\cite{WitrisalSPM2016Copy} and is used for the simulations in Sec.~\ref{sec:results}. Consider that assuming the statistical model to be correct, the distance variance will attain the \ac{crlb} for a single-distance measurement, as the \ac{nlos} measurements are taken into account by the data association algorithm. 
%

%
We define the nested vectors $\bm{z}^{(j)}_{n} = [{\bm{z}^{(j)\text{T}}_{n,1}} \, ... \, {\bm{z}^{(j)\text{T}}_{n,M_n^{(j)}}}]^\text{T}$ and $\bm{z}_n = [\bm{z}^{(1)\, \text{T}}_{n} ... \, \bm{z}^{(J)\,\text{T}}_{n}]^\text{T}$, where the latter denotes the joint measurement vector per time $n$. All of its components are used as noisy ``measurements'' by the proposed algorithm.

\section{System Model} \label{sec:system_model}
We consider a mobile agent to be moving along an unknown trajectory as depicted in Fig.~\ref{fig:eye_catcher}. The current state of the agent is described by the state vector $\bm{x}_n = [\bm{p}_n ^\text{T}\; \bm{v}_n^\text{T} ]^\text{T}$, which is composed of the mobile agent's position $\bm{p}_n = [p_{\text{x}\s n}\; p_{\text{y}\s n}]^\text{T}$ and velocity $\bm{v}_n = [v_{\text{x}\s n}\; v_{\text{y}\s n}]^\text{T}$.
The state evolves over time $n$ according to a predefined state transition \ac{pdf} $\Upsilon(\bm{x}_n|\bm{x}_{n-1})$, where the usual first-order Markov assumption is applied\cite{ArulampalamTSP2002}.
\subsection{Data Association Model} \label{sec:association_model}

At each time $n$ and for each anchor $j$, the measurements, i.e., the components of $\bm{z}_n^{(j)}$ are subject to data association uncertainty. Thus, it is not known which measurement $\bm{z}_{n,m}^{(j)}$ originated from the \ac{los}, or which one is due to an \ac{mpc}. It is also possible that a measurement $\bm{z}_{n,m}^{(j)}$ did not originate from any physical component, but from \acp{fa} of the prior channel estimation and detection algorithm. Our model only distinguishes between ``LOS measurements'' originating from the \ac{los} and ``\ac{nlos} measurements'', i.e., measurements due to \acp{mpc} or \acp{fa}. Based on the concept of \ac{pda} \cite{BarShalom1995}, we define an association variable
\begin{equation}\label{eq:association}
 a^{(j)}_{n} \rmv \rmv = \rmv 
 \begin{cases} 
 m \rmv\rmv \in \rmv\rmv \mathcal{M}_n^{(j)}  \rmv\rmv \rmv\rmv, &
 \begin{minipage}[t]{53mm}$\bm{z}_{n,m}^{(j)}$ is the LOS measurement in $\bm{z}_n^{(j)}$ \end{minipage}\\
   0 \, , & \begin{minipage}[t]{53mm} there is no LOS measurement in $\bm{z}_n^{(j)}$ \end{minipage}
  \end{cases}
\end{equation}
Assuming the number of \ac{nlos} measurements to follow a uniform distribution (so called ``non-parametric model''), the joint \ac{pmf} of $a^{(j)}_{n}$ and $M_n^{(j)}$ can be shown to be proportional to the function\cite{BarShalom1995}
\begin{equation}  \label{eq:prior_association}
 h(a^{(j)}_{n}\rmv\rmv\rmv, M_n^{(j)} ;u_n^{(j)}\rmv\rmv\rmv,q_n^{(j)}) =
  \begin{cases} 
 \frac{\pe}{M_n^{(j)}} \, , \rmv\rmv&  a^{(j)}_{n} \in \mathcal{M}_n^{(j)} \\[2mm] 1\minus\pe\,, \rmv\rmv\rmv & a^{(j)}_{n} = 0
  \end{cases}
\end{equation}
where $\pe$ is the probability that there is a \ac{los} measurement for the current set of measurements defined in Sec. \ref{sec:pnl} and $u_n^{(j)}$ and $q_n^{(j)}$ are defined in Sections \ref{sec:amplitude_model} and \ref{sec:pnl}, respectively.
We also define the joint vectors $\bm{a}_n = [a^{(1)}_{n}\, ... \, a^{(J)}_{n}]^\text{T}$ and $\bm{M}_n = [M_n^{(1)}\, ... \, M_n^{(J)}]^\text{T}$.

\subsection{Delay/Distance Model} \label{sec:delay_model}
To incorporate the data association procedure into the measurement process we define two \acp{lhf}, one for the \ac{los} event and one for the \ac{nlos} event.

Let the range-only measurement vector be ${\tilde{\bm{z}}^{(j)}_{n,m}} = [\hat{d}_{n,m}^{(j)}~ \sigmadhat]^\text{T}$.
First we define the \ac{los} \ac{lhf} as
\begin{equation}
 f_\text{L}({\tilde{\bm{z}}^{(j)}_{n,m}} | \bm{p}_n) =\mathcal{N}(\hat{d}^{(j)}_{n,m};\, \dlos,\, \sigmadhat)
\end{equation}
where $\mathcal{N}(\cdot)$ denotes a Gaussian \ac{pdf} of the \ac{rv} $\hat{d}^{(j)}_{n,m}$ with mean $\dlos$ and standard deviation $\sigmadhat$.
 The \ac{los} distance is geometrically related to the agent position via 
\begin{equation}\label{eq:multilateration}
d^{(j)}_{\text{LOS}\s n}(\bm{p}_n) = \norm{\bm{p}_n - \bm{p}_\text{A}^{(j)}}{}.
\end{equation}
Next, we define the \ac{nlos} \ac{lhf} as\cite{Yu2020}
\begin{equation}  \label{eq:nlos_lhf}
 f_\text{NL}(\hat{d}_{n,m}^{(j)} | \bm{p}_n ) \rmv=\rmv  P_\text{MP} \,f_\text{MP}(\hat{d}_{n,m}^{(j)} | \bm{p}_n)+ (1 \minus P_\text{MP}) \, f_\text{FA}(\hat{d}_{n,m}^{(j)}),
\end{equation}
which represents a weighted sum of two \acp{lhf}, with $P_\text{MP}$ acting as a weighting coefficient: The \ac{fa} \ac{lhf} $f_\text{FA}(\hat{d}_{n,m}^{(j)}) = \mathcal{U}(0,d_\text{max})$, which is a uniform distribution with the maximum distance $d_\text{max}$ and the multipath \ac{lhf}
\begin{equation} \label{eq:multipath_lhf}
f_\text{MP}( \hat{d}_{n,m}^{(j)} | \bm{p}_n) = 
  \begin{cases} 
\frac{\gamma_\text{f}+\gamma_\text{r}}{\gamma_\text{f}^{2}} (1 \minus e^{-\frac{\Delta^{(j)}_{n, m}}{\gamma_\text{r}}}) e^{-\frac{\Delta^{(j)}_{n, m}}{\gamma_\text{f}}}\, , &  \Delta^{(j)}_{n, m}>0 \\
 0 \,, & \Delta^{(j)}_{n, m} \leq 0
  \end{cases}
\end{equation} 
which is a double exponential function\cite{Karedal2007} with the distance difference $\Delta^{(j)}_{n, m} = \hat{d}^{(j)}_{n, m} - \dlos - B $. $\gamma_r$ is the rise distance, $\gamma_f$ the fall distance, and $B$ is a bias value. 

Finally, we define the overall-distance \ac{lhf} as 
\begin{equation} \label{eq:single_delay_like}
f({\tilde{\bm{z}}^{(j)}_{n,m}} | \bm{p}_n ,  a^{(j)}_{n}) =
  \begin{cases} 
f_\text{L}({\tilde{\bm{z}}^{(j)}_{n,m}}  | \bm{p}_n) \, , &  a^{(j)}_{n} = m \\
  f_\text{NL}(\hat{d}_{n,m}^{(j)} | \bm{p}_n) \,, & a^{(j)}_{n} \neq m
  \end{cases}\, .
\end{equation}
The shape of \eqref{eq:single_delay_like} is depicted in Fig.~\ref{fig:single_delay_like}.

\subsection{Amplitude Model} \label{sec:amplitude_model}

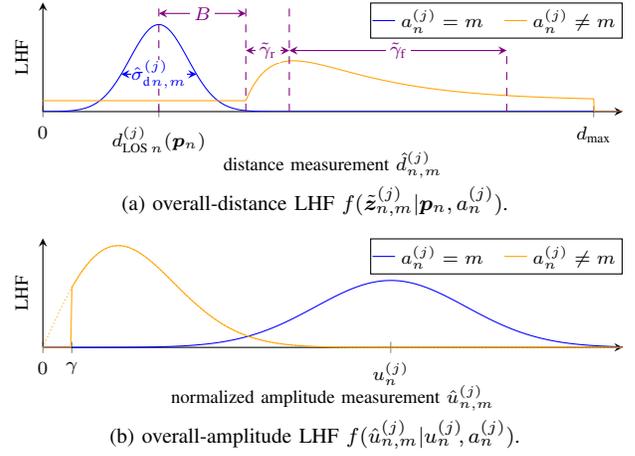
\begin{figure}[t]
\centering
\setlength{\figurewidth}{0.425\textwidth}
\setlength{\figureheight}{0.08\textwidth}

 \setlength{\belowcaptionskip}{0pt}
\captionsetup[subfloat]{farskip=8pt,captionskip=0pt} 
\subfloat[overall-distance \ac{lhf}  $f({\tilde{\bm{z}}^{(j)}_{n,m}} | \bm{p}_n ,  a^{(j)}_{n})$.\label{fig:single_delay_like}]{
%
%

\begin{tikzpicture}[
declare function={
func(\x)=(\x>0.8)*(0.6*\x)+and(\x>0.6,\x<=0.8)*(2/3*\x)+(\x<=0.6)*(\x*0.75);
pdp(\tau,\Om,\gl,\gr,\chi,\taumin)=(\tau>\taumin)*\Om * ( \gl + \gr )/( \gl*( \gl + \gr*( 1 - \chi ))) * exp((-\tau+\taumin)/\gl) * (1-\chi*exp((-\tau+\taumin)/\gr));
nlos(\x) = (\x<19)*(0.95*pdp(\x,1,3,1.5,1,7)+0.05);
sq(\y,\p)=pow(\y,\p);
gauss(\x,\mu,\sigma) = 1/(\sigma*sqrt(2*pi))*exp(-((\x-\mu)^2)/(2*\sigma^2));
},
myarrow/.style= {arrows={{stealth[length=1mm,width=1.2mm]}-{stealth[length=1mm,width=1.2mm]}}, thin}
]


\begin{axis}[%
simple style,
xmin=0,
xmax=20,
xtick={0,4,19},
xticklabels={{$0$},{$d^{(j)}_{\text{LOS}\,n}(\bm{p}_n)$},{$d_\text{max}$}},
xlabel={distance measurement $\hat{d}_{n,m}^{(j)}$},
xlabel style={yshift=2.5mm},
ylabel={LHF},
ymin=0,
ymax=0.5,
ytick={1},
yticklabels={{$\alpha_0$}},
axis x line*=bottom,
axis y line*=left,
axis lines=left,
legend style={at={(1, 1)},
inner sep = 0.5pt, legend cell align=left, legend columns = 4, /tikz/every even column/.append style={column sep= 2mm}, font=\scriptsize},
legend image post style={xscale=0.5, yscale=0.5},
samples=1000,domain=0:20,
]


\addplot [color=mycolor01
, mark options={solid, mycolor03}]
 {gauss(x,4,1)};
\addlegendentry{$a^{(j)}_{n} = m$}

\addplot [color=mycolor06
]
{nlos(x)};
\addlegendentry{$a^{(j)}_{n} \neq m$}



\addplot[forget plot,ycomb, color=mycolor03, dashed, mark=none, mark options={solid, mycolor03, scale = 1}] coordinates { (4,0.5) };
\addplot[forget plot,ycomb, color=mycolor03, dashed, mark=none, mark options={solid, mycolor03, scale = 1}] coordinates { (7,0.5) };
\addplot[forget plot,ycomb, color=mycolor03, dashed, mark=none, mark options={solid, mycolor03, scale = 1}] coordinates { (8.5,0.5) };
\addplot[forget plot,ycomb, color=mycolor03, dashed, mark=none, mark options={solid, mycolor03, scale = 1}] coordinates { (7+3*3,0.28) };

\draw[myarrow, color=mycolor01] (axis cs: 2.7, {gauss(2.7,4,1)}) -- (axis cs: 5.3, {gauss(5.3,4,1)})
node[midway, fill=white, inner sep=0mm] {\scriptsize $\sigmadhat$};
\draw[myarrow, color=mycolor03] (axis cs: 4, 0.45) -- (axis cs: 7, 0.45)
node[midway, fill=white, inner sep=0.5mm] {\scriptsize $B$};
\draw[myarrow, color=mycolor03] (axis cs: 7, 0.28) -- (axis cs: 8.5, 0.28)
node[midway, fill=white, inner sep=0.5mm] {\scriptsize $\tilde{\gamma}_\text{r}$};
\draw[myarrow, color=mycolor03] (axis cs: 8.5, 0.28) -- (axis cs: 7+3*3, 0.28)
node[midway, fill=white, inner sep=0.5mm] {\scriptsize $\tilde{\gamma}_\text{f}$};

\end{axis}
\end{tikzpicture}%
}
 
 \subfloat[overall-amplitude \ac{lhf}  $f(\hat{u}^{(j)}_{n,m} | u_n^{(j)} \rmv\rmv  , a^{(j)}_{n})$.\label{fig:single_amplitude_like}]{
%
%

\begin{tikzpicture}[
declare function={
func(\x)=(\x>0.8)*(0.6*\x)+and(\x>0.6,\x<=0.8)*(2/3*\x)+(\x<=0.6)*(\x*0.75);
rice(\x,\mu,\sigma) = \x/(\sigma^2)*exp(-(\x^2+\mu^2)/(2*\sigma^2));
trice(\x,\mu,\sigma,\tmin) = (\x>\tmin)*rice(\x,\mu,\sigma);
gauss(\x,\mu,\sigma) = 1/(\sigma*sqrt(2*pi))*exp(-((\x-\mu)^2)/(2*\sigma^2));
tgauss(\x,\mu,\sigma,\tmin) = (\x>\tmin)*gauss(\x,\mu,\sigma);
},
myarrow/.style= {arrows={{stealth[length=1mm,width=1.2mm]}-{stealth[length=1mm,width=1.2mm]}}, thin}
]


\begin{axis}[%
simple style,
xmin=0,
xmax=10,
xtick={0,0.5,6},
xticklabels={{$0$},{$\gamma$},{$u_n^{(j)}$}},
xlabel={normalized amplitude measurement $\hat{u}_{n,m}^{(j)}$},
xlabel style={yshift=2.5mm},
ylabel={LHF},
ymin=0,
ymax=0.5,
ytick={1},
yticklabels={{$\alpha_0$}},
axis x line*=bottom,
axis y line*=left,
axis lines=left,
legend style={at={(1, 1)},
legend cell align=left, legend columns = 4, /tikz/every even column/.append style={column sep= 2mm}, font=\scriptsize, inner sep = 0.5pt},
legend image post style={xscale=0.5, yscale=0.5},
samples=1000,domain=0:20,
]


\addplot [color=mycolor01
]
{tgauss(x,6,1.3,0.5)};
\addlegendentry{$a^{(j)}_{n} = m$}

\addplot [color=mycolor06]
 {trice(x,0,1.3,0.5)};
\addlegendentry{$a^{(j)}_{n} \neq m$}

\addplot [color=mycolor06, densely dotted, forget plot]
 {trice(x,0,1.3,0)};


\end{axis}
\end{tikzpicture}%
}

 \caption{Graphical representation of the stochastic models constituting the overall \ac{lhf} for a single measurement.}
 \label{fig:single_measurment_like}
 \vspace{-2mm}
\end{figure}

As for the delay model in Sec.~\ref{sec:delay_model}, we start by defining the \ac{los} amplitude \ac{lhf} as
$ f_\text{L}(\hat{u}^{(j)}_{n, m}| u^{(j)}_{n}) \propto f_\text{Rice}(\hat{u}^{(j)}_{n, m};1,u^{(j)}_{n}) \,\iota(\hat{u}^{(j)}_{n, m} - \gamma)$ and the \ac{nlos} amplitude \ac{lhf} as $f_\text{NL}(\hat{u}^{(j)}_{n, m}) \propto f_\text{Rayl}(\hat{u}^{(j)}_{n, m}; 1) \,\iota(\hat{u}^{(j)}_{n, m} - \gamma)$, where $\iota(\hat{u}^{(j)}_{n, m} - \gamma)$ is the unit step function with threshold $\gamma$ that truncates the distributions. The \ac{pdf} $f_\text{Rice}$ is a Rician distribution and \ac{pdf} $f_\text{Rayl}$ is a Rayleigh distribution with non-centrality parameter $u^{(j)}_{n}$ and respective spread parameters equal to $1$. Note that due to truncation the functions need to be scaled to represent proper \acp{pdf} \cite{LerroACC1990}.
The overall-amplitude \ac{lhf} is given by
 \vspace{-0.8mm}
\begin{equation} \label{eq:single_ampl_like}  \vspace{-0.3mm}
f(\hat{u}^{(j)}_{n, m} | u^{(j)}_{n} ,  a^{(j)}_{n}) =
  \begin{cases} 
f_\text{L}(\hat{u}^{(j)}_{n, m} | u^{(j)}_{n}) \, , &  a^{(j)}_{n} = m \\
  f_\text{NL}(\hat{u}^{(j)}_{n, m}) \,, & a^{(j)}_{n} \neq m 
  \end{cases}
\end{equation}
which is shown in Fig.~\ref{fig:single_amplitude_like}. This model represents the distribution of amplitude estimates of a single complex baseband signal in additive Gaussian noise obtained using maximum likelihood estimation and generalized likelihood ratio test detection \cite{VenusICC2020,Kay1993, Kay1998}. Consider that for the model to be true, the \acp{mpc} in \eqref{eq:signal_model}, i.e., all components except for the \ac{los}, have to be represented by a stochastic process with zero mean (i.e. dense multipath component model\cite{WitrisalSPM2016Copy}).
The above definition is similar to \cite{LerroACC1990}. However, we use the normalized amplitude $u^{(j)}_{n}$ \cite{LeitingerICC2019}, i.e., the spreading parameters of the Rayleigh and Rician distribution, which constitute the overall-amplitude \ac{lhf} \eqref{eq:single_ampl_like}, are equal to $1$, avoiding the necessity to track these parameters.
We model the temporal evolution of $u^{(j)}_{n}$ as a first order Markov process, which is defined by a state transition \ac{pdf} $\Phi(u_n^{(j)}|u_{n-1}^{(j)})$. The amplitudes of all anchors $j$ are assumed to be independent stochastic processes, ignoring possible geometric information available, as for channels with strong multipath propagation received signal strength measurements tend to be error prone.
%
%
 %
%
We also define the joint amplitude vector $\bm{u}_n = [u^{(1)}_{n}\, ... \, u^{(J)}_{n}]^\text{T}$.

\subsection{LOS Existence Probability Model}\label{sec:pnl}

We model the \ac{los} existence probability given in \eqref{eq:prior_association} as
$\pe = \pd\, q_n^{(j)}$. 
The so-called probability of detection $\pd$ is modelled according to Sec. \ref{sec:amplitude_model} by assuming that the proposed algorithm is applied after a generalized likelihood ratio test detector. 
%
%
That is, $\pd$ is completely determined by the normalized amplitude $u_n^{(j)}$ and $\gamma$, which represents the detection threshold and is a constant to be chosen.
%
%
%
%
$q_n^{(j)}$ is the probability of the event that the \ac{los} is \textit{not} obstructed, which is referred to as \ac{los} probability in the following, and acts as a prior probability to the detection event.
According to \cite{PapaICASSP2015,SoldiTSP2019}
, we model $q_n^{(j)}$ as discrete \ac{rv} that takes its values from a finite set $\mathcal{Q} = \{\omega_1,\, ... \,, \omega_Q\}$, where $\omega_i \in (0,1]$. The temporal evolution of $q_n^{(j)}$ is modelled by a first-order Markov process, which results in a conventional Markov chain, with $[\bm{Q}^{(j)}]_{i,k} = \Psi(q_n^{(j)} = \omega_i| q_{n \minus 1}^{(j)} = \omega_k$) being the elements of the transition matrix.
The \ac{los} probabilities for different sensors $j$ are assumed to be independent. 
We also define the joint \ac{los} probability vector $\bm{q}_n = [q^{(1)}_{n}\, ... \, q^{(J)}_{n}]^\text{T}$.

\subsection{Joint Measurement Likelihood Function}

Under commonly used assumptions about the statistics of the measurements\cite{MeyerProc2018}, the joint \ac{lhf} for all measurements per anchor $j$ and time $n$ can be written as
\vspace{-2mm}
\begin{equation}
f(\bm{z}_n^{(j)} | \bm{p}_n ,\rmv\rmv u^{(j)}_{n} \rmv\rmv\rmv,\rmv  a^{(j)}_{n}) \rmv = \rmv\rmv\rmv\rmv \rmv  \prod_{m=1}^{\,\,M_{n}^{(j)}} \rmv\rmv\rmv f({\tilde{\bm{z}}^{(j)}_{n,m}}| \bm{p}_n , \rmv  a^{(j)}_{n}) \s f(\hat{u}^{(j)}_{n, m} | u^{(j)}_{n} \rmv\rmv\rmv,\rmv  a^{(j)}_{n})
\end{equation}
By neglecting all constant terms, we define the pseudo \ac{lhf}
\begin{align}\label{eq:likelihood}
&g( \bm{z}_{n}^{(j)} ; \bm{p}_n, u^{(j)}_{n} ,  a^{(j)}_{n})  \nonumber  \\ &
= \prod_{m=1}^{M_{n}^{(j)}} f_\text{NL}(\hat{d}_{n,m}^{(j)} | \bm{p}_n) \times \rmv\rmv
  \begin{cases} 
1  , &  a^{(j)}_{n} = 0 \\
 \Lambda( \bm{z}_{n, a^{(j)}_{n}}^{(j)} | \bm{p}_n, u^{(j)}_{n} )  , & a^{(j)}_{n} \in \mathcal{M}_n^{(j)}
  \end{cases} \\[-7mm] \nonumber
\end{align} 
where
\vspace{-3mm}
\begin{equation} \label{eq:likelihood_ratio}
  \Lambda( \bm{z}_{n,m}^{(j)} | \bm{p}_n, u^{(j)}_{n} ) = \frac{f_\text{L}(\tilde{\bm{z}}^{(j)}_{n, m}  | \bm{p}_n) \, f_\text{L}(\hat{u}^{(j)}_{n, m} | u^{(j)}_{n})}{f_\text{NL}(\tilde{\bm{z}}^{(j)}_{n, m}  | \bm{p}_n)\, f_\text{NL}(\hat{u}^{(j)}_{n, m}) }
\end{equation}
is the likelihood ratio.
Note that 
the product of all \ac{nlos} events in \eqref{eq:likelihood} is not a constant and thus cannot be neglected.

\subsection{Joint Posterior and Factor Graph}  \label{sec:factor_graph}

\begin{figure}[t]
 \centering
 %
 %
 \includegraphics{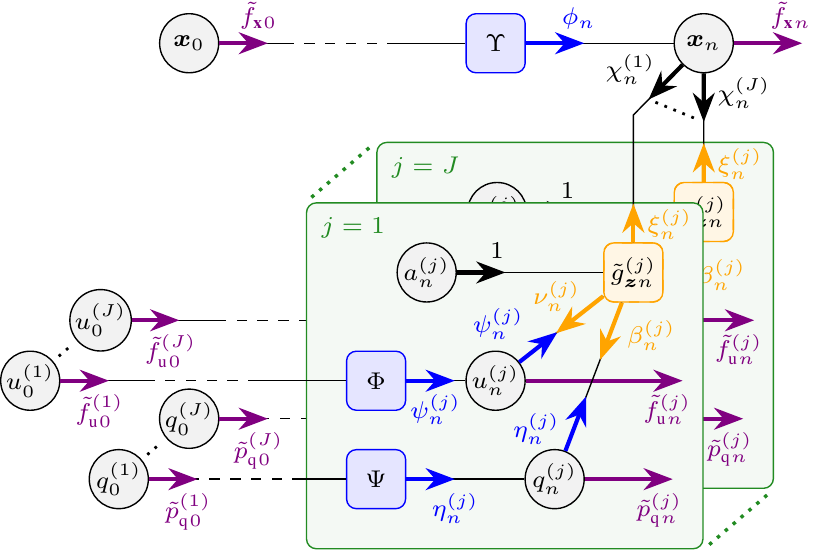}
 \vspace*{-2.5mm}
 \caption{Factor graph representing the factorization of the joint posterior \ac{pdf} in \eqref{eq:factorization1} and the messages according to the SPA (see Sec.~\ref{sec:spa}).}\label{fig:factor_graph}
 \vspace{-3mm}
\end{figure}

Let 
$\bm{z} = [\bm{z}^\text{T}_{1}\, ... \, \bm{z}^\text{T}_{n}]^\text{T}$, 
$\bm{x} = [\bm{x}^\text{T}_{1}\, ... \, \bm{x}^\text{T}_{n}]^\text{T}$, 
$\bm{a} = [\bm{a}^\text{T}_{1}\, ... \, \bm{a}^\text{T}_{n}]^\text{T}$, 
$\bm{u} = [\bm{u}^\text{T}_{1}\, ... \, \bm{u}^\text{T}_{n}]^\text{T}$, 
$\bm{q} = [\bm{q}^\text{T}_{1}\, ... \, \bm{q}^\text{T}_{n}]^\text{T}$, 
and $\bm{M} = [\bm{M}^\text{T}_{1}\, ... \, \bm{M}^\text{T}_{n}]^\text{T}$.
Applying Bayes' rule as well as some commonly used independence assumptions\cite{MeyerProc2018,LeitingerGNSS2016} the joint posterior for all states up to time $n$ and all $J$ anchors, can be derived up to a constant factor as 
\begin{align} \label{eq:factorization1}
&  f(\bm{x}, \rmv\bm{a}, \rmv\bm{u}, \rmv\bm{q}, \rmv\bm{M} | \bm{z} ) 
\nonumber\\
 &\propto  f(\bm{z} |  \bm{x},\rmv\bm{a}, \rmv\bm{u}, \rmv\bm{q} ) \, f( \bm{x},\rmv\bm{a}, \rmv\bm{u}, \rmv\bm{q}) \nonumber\\
& =  f(\bm{z} |  \bm{x},\rmv\bm{a}, \rmv\bm{u}, \rmv\bm{q} ) \, f(\rmv\bm{a} | \bm{u}, \rmv \bm{q})\, f(\rmv \bm{x}) \, p(\rmv\bm{q}) \, f(\rmv \bm{u})\nonumber \\
&\propto f( \bm{x}_0) \rmv\rmv \prod^{J}_{j= 1} p( q_0^{(j)}) \,f( u_0^{(j)})  \rmv
  \rmv\rmv\prod^{n}_{n'= 1}  \rmv\rmv\rmv \Upsilon( \bm{x}_{n'} | \bm{x}_{n'\minus1})\, \Phi( u_{n'}^{(j)} | u_{n'\minus1}^{(j)}) \nonumber \\& ~~~~~\times \Psi( q_{n'}^{(j)} | q_{n'\minus1}^{(j)}) \,  \tilde{g}( \bm{z}_{n'}^{(j)} ; \bm{p}_{n'}, u^{(j)}_{n'} ,  a^{(j)}_{n'},  q^{(j)}_{n'})\,,
\end{align}
with 
$\tilde{g}(\bm{z}_{n}^{(j)}; \bm{p}_{n},\rmv\rmv u^{(j)}_{n} \rmv\rmv\rmv\rmv\rmv, \rmv a^{(j)}_{n}\rmv\rmv\rmv\rmv, \rmv q_n^{(j)})
\rmv\rmv = \rmv\rmv 
h(a_{n}^{(j)} ; u_n^{(j)},q_n^{(j)})\,
g( \bm{z}_{n}^{(j)} ; \bm{p}_n, $ $\rmv u^{(j)}_{n}\rmv\rmv\rmv\rmv\rmv , \rmv a^{(j)}_{n})$. 
For the sake of brevity, we refer to this expression as  $\tilde{g}_{\bm{z} n}^{(j)}(\cdot)$ in the rest of the work.
Note that $\bm{M}$ is fixed and thus constant, as it is defined implicitly by the measurements $\bm{z}$.
%
This factorization of the joint posterior \ac{pdf}
can be visually represented by the factor graph shown in Fig.~\ref{fig:factor_graph}.
Further note that \eqref{eq:factorization1} is a mixture of discrete \acp{pmf} and continuous \acp{pdf}.

\section{Algorithm} \label{sec:algorithm}

%

\subsection{Problem Statement}  

Our goal is to estimate the agent state $\bm{x}_n$. This can be done by calculating the \ac{mmse} \cite{Kay1993}
\begin{equation}\label{eq:mmse}
\hat{\bm{x}}^\text{MMSE}_{n} \,\triangleq \int \rmv \bm{x}_{n} \, f(\bm{x}_{n} | \bm{z} )\, \mathrm{d}\bm{x}_{n} \,.
\end{equation}
with $\hat{\bm{x}}^{\text{MMSE}}_n = [\hat{\bm{p}}^{\text{MMSE T}}_n \, \hat{\bm{v}}^{\text{MMSE T}}_n ]^\text{T}$. Furthermore, we also calculate 
\begin{align}
\hat{u}^{(j)\s\text{MMSE}}_{n} &\,\triangleq \int \rmv u_n^{(j)} \, f(u_n^{(j)} | \bm{z} )\, \mathrm{d} u_n^{(j)}, \label{eq:mmseAmpl}\\
\hat{q}^{(j)\s\text{MMSE}}_{n} &\,\triangleq \sum_{\omega_i \in \mathcal{Q}} \rmv \omega_i \, p(q_n^{(j)} = \omega_i | \bm{z} )
\,.\label{eq:mmseplos}
\end{align}
In order to obtain \eqref{eq:mmse}, \eqref{eq:mmseAmpl}, and \eqref{eq:mmseplos}, marginalization of  the joint posterior has to be performed. In general this is computationally infeasible\cite{MeyerProc2018}. To counteract this problem, we use a \ac{spa} based algorithm introduced in the next section.
 
 \subsection{Marginal Posterior and Sum-Product Algorithm (SPA)}  \label{sec:spa}
The marginal posterior can be calculated efficiently by passing messages on the factor graph according to the \ac{spa}\cite{KschischangTIT2001}.
 The presented algorithm is an adaptation of the algorithms presented in \cite{MeyerProc2018, LeitingerICC2017} to the factor graph shown in Fig.~\ref{fig:factor_graph}.
As the filter shall be executable online, we only pass messages forward in time. This makes the factor graph in Fig.~\ref{fig:factor_graph} an acyclic graph. For acyclic graphs the \ac{spa}
yields \textit{exact results} for the marginal posterior \cite{KschischangTIT2001}.
%
%
%
%
At time $n$, the following calculations are performed for all $J$ anchors; We start by defining the prediction messages, where $\tilde{f}_{\text{\textbf{x}}\s n\minus 1}(\cdot)$, $\tilde{f}_{\text{u}\s n\minus 1}^{(j)}(\cdot)$ and $\tilde{p}_{\text{q}\s n\minus 1}^{(j)}(\cdot)$ are messages of the previous time $n\minus 1$, as 
\vspace{-1mm}
\begin{align}\vspace{-1mm} \label{eq:message1}
 \phi_n( \bm{x}_n) &= \int {\Upsilon}( \bm{x}_n | \bm{x}_{n\minus 1})\, \tilde{f}_{\text{\textbf{x}} n\minus 1}(\bm{x}_{n\minus 1})\,\mathrm{d}\bm{x}_{n\minus 1}\,,\\[0pt]
 \psi_n^{(j)}( u_n^{(j)}) &= \int \Phi( u_n^{(j)} | u_{n\minus1}^{(j)})\, \tilde{f}_{\text{u}n\minus 1}^{(j)}(u_{n\minus 1}^{(j)})\,\mathrm{d}u_{n\minus 1}^{(j)}\,,\\[-2pt]
%
 \eta_n^{(j)}(q_n^{(j)}) &= \sum_{q_{n\minus 1}^{(j)}=1}^{N_q} \Psi( q_n^{(j)} | q_{n\minus 1}^{(j)})\, \tilde{p}_{\text{q}n\minus 1}^{(j)}(q_{n\minus 1}^{(j)}).
\end{align}
Next, we define the measurement update messages as
\begin{align}
\xi_n^{(j)}(\bm{x}_n) &= \rmv\rmv\rmv\rmv \int \rmv\rmv\rmv\rmv \psi_n^{(j)}(u_n^{(j)}) \rmv\rmv\rmv\rmv\rmv\rmv\rmv  \sum_{q_{n}^{(j)}=1}^{N_q} \rmv\rmv\rmv\rmv\rmv\rmv \eta_n^{(j)}\rmv (q_n^{(j)}) \rmv\rmv\rmv\rmv\rmv\rmv\rmv\rmv   \sum_{a_{n}^{(j)}=1}^{M_n^{(j)}} \rmv\rmv\rmv\rmv\tilde{g}_{\bm{z} n}^{(j)}(\cdot)  \,\mathrm{d} u_{n}^{(j)}, \\[-2pt]
%
%
\chi_n^{(j)}(\bm{x}_n) &= \phi_n( \bm{x}_n) \prod_{j^{\prime}=1}^{J}  \xi_n^{(j^{\prime})}(\bm{x}_n) /  \xi_n^{(j)}(\bm{x}_n),\\[-2pt]
 \nu_n^{(j)}(u_n^{(j)}) &= \rmv\rmv\rmv\rmv\rmv \rmv\rmv  \sum_{q_{n}^{(j)}=1}^{N_q} \rmv\rmv\rmv\rmv \eta_n^{(j)}(q_n^{(j)}) \rmv\rmv \rmv \int \rmv \rmv\rmv \chi_n^{(j)}(\bm{x}_n) \rmv\rmv  \rmv\rmv \rmv\rmv \sum_{a_{n}^{(j)}=1}^{M_n^{(j)}} \tilde{g}_{\bm{z} n}^{(j)}(\cdot)  \,\mathrm{d}\bm{x}_{n},\\[-2pt]
\label{eq:mesageend}
 \beta_n^{(j)}(q_n^{(j)}) &= \rmv\rmv\rmv\rmv \int\rmv\rmv\rmv\rmv\rmv\rmv\rmv\rmv \int \rmv\rmv\rmv \psi_n^{(j)}( u_n^{(j)}) \,\chi_n^{(j)}(\bm{x}_n)\rmv\rmv\rmv\rmv\rmv\sum_{a_{n}^{(j)}=1}^{M_n^{(j)}}\rmv\rmv\rmv \tilde{g}_{\bm{z} n}^{(j)}(\cdot)  \,\mathrm{d}\bm{x}_{n} \,\mathrm{d} u_{n}^{(j)}.
\end{align}
Finally, we calculate the posterior distributions as
$f(\bm{x}_n| \bm{z}) \propto  \tilde{f}_{\text{\textbf{x}}\s n}(\bm{x}_n) \rmv\rmv =  \rmv\rmv \phi_n( \bm{x}_n) \prod_{j=1}^{J} \xi_n^{(j)}(\bm{x}_n)$, $f(u_n^{(j)}| \bm{z}) \propto \tilde{f}_{\text{u}\s n}^{(j)}(u_n^{(j)}) \rmv\rmv =  \psi_n^{(j)}(u_n^{(j)}) \, \nu_n^{(j)}(u_n^{(j)})$
and 
$p(q_n^{(j)}| \bm{z}) \propto \tilde{p}_{\text{q}\s n}^{(j)}(q_n^{(j)}) =  \eta_n^{(j)}(q_n^{(j)}) $ $ \times \beta_n^{(j)}(q_n^{(j)})$.
%
%
%
Since a direct calculation of the integrals in equations \eqref{eq:message1}-\eqref{eq:mesageend} is intractable, a particle-based approximation\cite{ArulampalamTSP2002} is used. See \cite{LeitingerICC2017} and \cite{MeyerProc2018} for details.

 \subsection{Initialization}\label{sec:init}

We propose to initialize the normalized amplitude \acp{pdf} as $\tilde{f}_{\text{u}\s 0}^{(j)}(u_0^{(j)}) = \mathcal{U}(0,u_\text{max})$, where $u_\text{max}$ is a constant to be chosen according to hardware specific limitations. The \ac{los} \acp{pmf} are initialized at $q_0^{(j)}=1$. 
Regarding the agent state $\bm{x}_0$, we assume the velocity to be initialized at $\bm{v}_0 = \bm{0}$, as we do not know in which direction we are moving.
Since we cannot make any assumptions about the angle that the agent takes with respect to any of the anchors, it is reasonable to draw the positions $\bm{p}_0$ uniformly on two-dimensional discs around each anchor $j$, which are bounded by the maximum possible distance $d_\text{max}$ and a sample is drawn from each of the $J$ discs with equal probability.


\section{Computational Results}\label{sec:results}

 \begin{figure}[t!]

\centering
\setlength{\abovecaptionskip}{0pt}
\setlength{\belowcaptionskip}{0pt}
 
\setlength{\figurewidth}{0.25\textwidth}
\setlength{\figureheight}{0.25\textwidth}

\def\datapath{./figures/track}
\includegraphics{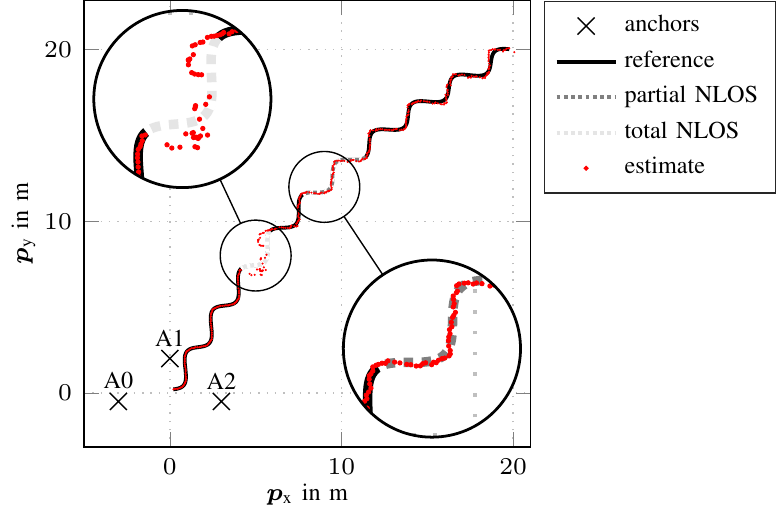}
\vspace{-1mm}
\caption{Simulated trajectory and anchor setup together with a single position estimate, corresponding to the measurement shown in Fig.~\ref{fig:meas_space}}\label{fig:track}
\vspace{-4.5mm}
\end{figure}

We evaluate the proposed algorithm using numerical simulation. To investigate the performance independently of the channel estimation and detection algorithm implementation and possible resulting artifacts, we directly generate the measurement vector $\bm{z}$ according to the system model in Sec.~\ref{sec:system_model}.

\subsection{Simulation Model}\label{sec:simulation_model}
\begin{figure}[t!]

\centering
\setlength{\abovecaptionskip}{0pt}
\setlength{\belowcaptionskip}{0pt}
 
\setlength{\figurewidth}{0.425\textwidth}
\setlength{\figureheight}{0.26\textwidth}

\vspace{-1mm}
\def\datapath{./figures/meas_space}
\def\pathnlos{./figures/pnlos}
\includegraphics{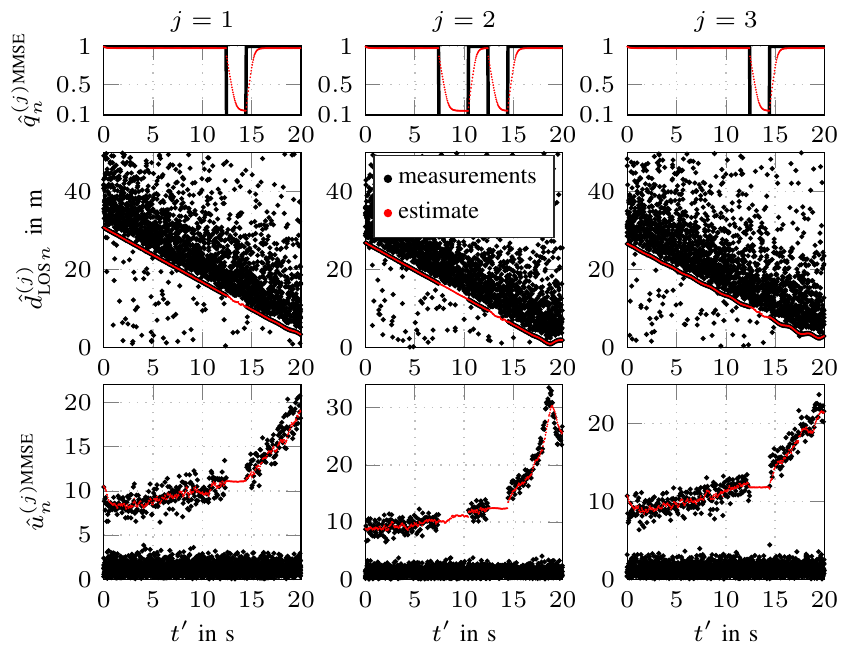}
\vspace{-1mm}
\caption{A single measurement realization and the respective estimates using the proposed algorithm (AL5). $\hat{d}^{(j)}_{\text{LOS}\s n}$ is calculated using \eqref{eq:multilateration} and \eqref{eq:mmse}.}\label{fig:meas_space}
\vspace*{-3mm}
\end{figure}

In the example scenario investigated, the agent moves along a curvy trajectory from a distant point to the centre of an object where three anchors are mounted (e.g. a car or a door). The trajectory is illustrated in Fig.~\ref{fig:track}. It is observed over a continuous measurement time $t^\prime \in [0, 20]\,\mathrm{s}$, with a constant sampling rate of $\Delta T = 50\,\mathrm{ms}$, resulting in $N=400$ discrete times steps $n \in \{1 \,...\,N\}$.  It comprises two \ac{olos} situations, a partial one, where only the \ac{los} to one anchor is blocked, and a full one, where the \ac{los} to all anchors is blocked. $\norm{\bm{v}_n}{}$ is set to vary around a velocity of $1.4 \,\mathrm{m/s}$. 
The normalized amplitudes are set to $\sqrt{ 30\,\mathrm{dB} }$ at $d^{(j)}_{\text{LOS}n} = 1\,\mathrm{m}$, with an exponential path-loss factor as low as $0.4$ to consider multipath propagation. We used an average rate of $10$ \ac{nlos} measurements per time $n$. The parameters of the \ac{nlos} \ac{lhf}, were set to $P_\text{MP} = 0.9$, $\gamma_\text{r}=1.5\,\mathrm{m}$, $\gamma_\text{f}=6\,\mathrm{m}$, $B= 0.2\,\mathrm{m}$ and $d_\text{max} = 50\,\text{m}$. Fig.~\ref{fig:meas_space} shows simulated measurements corresponding to a single realization of the trajectory.

\subsection{Inference Model}\label{sec:inference_model}
In the estimation algorithm, the agent motion, i.e. the state transition \ac{pdf} $\Upsilon(\bm{x}_n|\bm{x}_{n-1})$, is modelled by a linear, constant velocity and stochastic acceleration model\cite[p.~273]{BarShalom2002EstimationTracking}, i.e. $\bm{x}_n = \bm{A}\, \bm{x}_{n\minus 1} + \bm{B}\, \bm{w}_{n}$,
with
the acceleration process $\bm{w}_n$ being i.i.d. across $n$, zero mean, and Gaussian with covariance matrix ${\sigma_{\text{a}}}\, \bm{I}$, where $\bm{I}$ is a 2x2 identity matrix, ${\sigma_{\text{a}}} = 0.3\,\mathrm{m/s^2}$ is the acceleration standard deviation, and $\bm{A} \in \mathbb{R}^{\text{4x4}}$ and $\bm{B} \in \mathbb{R}^{\text{4x2}}$ are defined according to \cite[p.~273]{BarShalom2002EstimationTracking}, with $\Delta T$ as defined in Sec.~\ref{sec:simulation_model}.
The state transition \ac{pdf} of the normalized amplitudes is modelled as as Gaussian distribution $\Phi(u_n^{(j)}|u_{n-1}^{(j)}) $ $=\mathcal{N}(u_n^{(j)}; u_{n-1}^{(j)}, \sigma_u=0.2)$
, which is independent across $n$ and $j$. 
Thus, unlike the simulation model in Sec.~\ref{sec:simulation_model}, the amplitudes of all sensors $j$ are assumed to be independent (see Sec.~\ref{sec:amplitude_model}).
We use $\gamma = 0$, which is equivalent to using no detection threshold at all. Thus $ \pd = 1$, which leads to $\pe \equiv q_n^{(j)}$.
The set of possible \ac{los} probabilities is chosen as $\mathcal{Q} = \{0.1,\,0.2,\, ... \,, 1\}$. The state transition matrix $\bm{Q}^{(j)} = \bm{Q}$ is set as follows: $[\bm{Q}]_{1,1}=0.9$, $[\bm{Q}]_{10,10}=0.95$, $[\bm{Q}]_{2,1}=0.1$ and $[\bm{Q}]_{9,10}=0.05$. For $2\leq k \leq 9$, $[\bm{Q}]_{k,k} = 0.85$, $[\bm{Q}]_{k-1,k} = 0.05$ and $[\bm{Q}]_{k+1,k} = 0.1$. For all other tuples $\{i,k\}$, $[\bm{Q}]_{i,k} = 0$. 
We used $10^4$ particles for initialization and $10^3$ particles for inference during the track.

\subsection{Performance Results}
\begin{figure}[t]

\centering

\hspace*{0.95cm}
\includegraphics{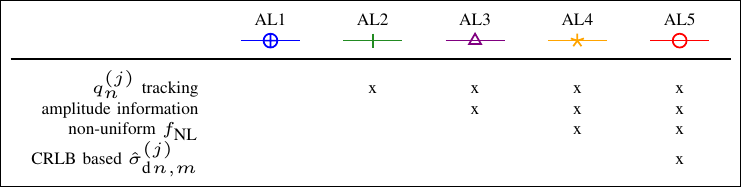}

\vspace{2mm}

\setlength{\abovecaptionskip}{0pt}
\setlength{\belowcaptionskip}{0pt}
 
\setlength{\figurewidth}{0.425\textwidth}
\setlength{\figureheight}{0.15\textwidth}

\def\datapath{./figures/mse_along_path}
\hspace*{-1.3mm}
\includegraphics{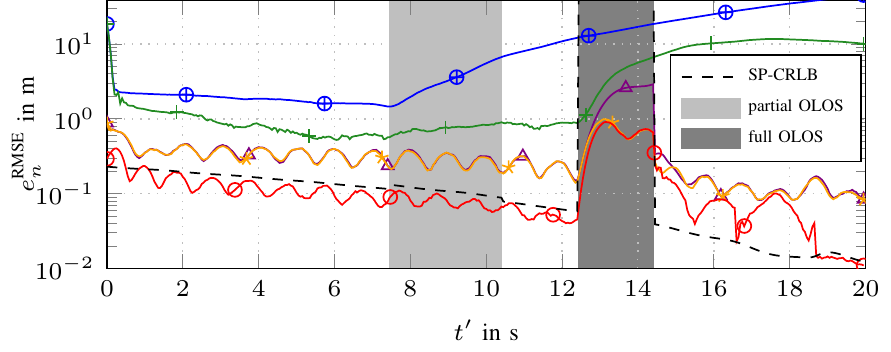}

\def\datapath{./figures/mse_cdf}
\includegraphics{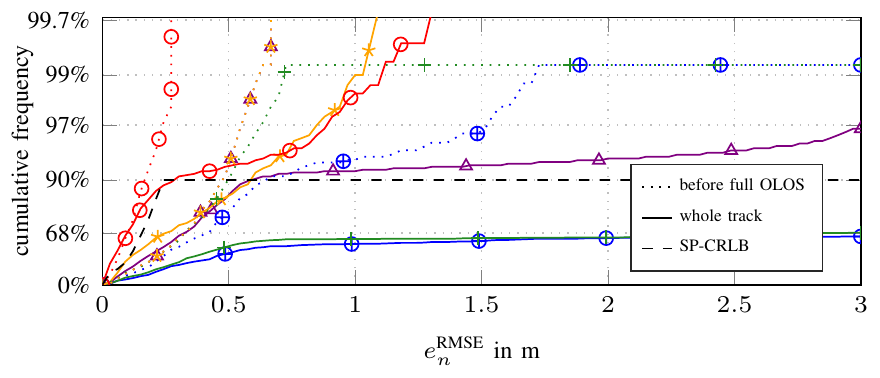}

\caption{Performance in terms of the \ac{rmse} of the estimated agent position determined using numerical simulation. The upper plot shows the estimated $e_{n}^{\text{RMSE}}$ as a function of the measurement time $t^\prime$. The lower plot shows the cumulative frequency of the $e_{n}^{\text{RMSE}}$ in inverse logarithmic scale.}\label{fig:results}

\end{figure}

We analyze the influence of the individual features of our algorithm with respect to the scenario described in Sec. \ref{sec:simulation_model}.
Fig.~\ref{fig:results} shows the algorithm variants implemented and the corresponding features that are enabled for an algorithm (x) or not ( ). 
When ``$q_n^{(j)}$ tracking'' is deactivated, we set $q_n^{(j)} = 0.999$ for all $n$, $j$. When we do not use ``\ac{crlb} based $\sigmadhat$'' measurements (see Sec. \ref{sec:channel_estimation}), it is set constant to $\sigmadhat=0.1\,\mathrm{m}$. Not applying the ``non-uniform $f_\text{NL}$'' means $P_\text{MP}=0$, and deactivating ``amplitude information'' means  $f_\text{L}(\hat{u}^{(j)}_{n, m} | u^{(j)}_{n}) / f_\text{NL}(\hat{u}^{(j)}_{n, m}) \triangleq 1$ in \eqref{eq:likelihood_ratio}.
All simulation results are shown in terms of the \ac{rmse} of the estimated agent position $e_{n}^{\text{RMSE}}~=~\sqrt{\E{\norm{\hat{\bm{p}}^{\text{MMSE}}_n -\bm{p}_n}{2}}}$, evaluated using a numerical simulation with 500 realizations. The \ac{rmse} is shown in two ways. First, as a function of the continuous measurement time $t^\prime$ and, second, as the cumulative frequency of the \ac{rmse} evaluated over the whole time span ($t^\prime \in [0, 20]\, \mathrm{s}$), as well as over the time span before the total \ac{nlos} situation ($t^\prime \in [0, 14.2]\, \mathrm{s}$).
As a performance benchmark we provide the \ac{crlb} for a single position measurement without tracking (SP-CRLB) \cite{Jourdan2008}.
Comparing the curves of Fig.~\ref{fig:results}, one can conclude that the \ac{rmse} is significantly lowered when additional features are activated. The RMSE of AL1, which represents a conventional multi-sensor \ac{pda}, is constantly above $2$ m. This is due to the large percentage of outliers, i.e., realizations where the algorithm completely loses the track. This is slightly improved by tracking $q_n^{(j)}$ (AL2), which leads to a reduced number of lost tracks. For AL3, we activate the amplitude information feature, which, in case of sufficient component SNR, significantly improves the performance as \ac{nlos} and \ac{los} measurements can be separated better. AL4 can additionally support the state estimation using \ac{nlos} measurements, as \eqref{eq:nlos_lhf} depends on the agent position $\bm{p}_n$ due to the non-uniform \ac{nlos} \ac{lhf}. 
This is especially beneficial in the full \ac{olos} situation as it significantly reduces the probability of a lost track. Finally we use \ac{crlb} based $\sigmadhat$ measurements for AL5. This additionally reduces the error, as the variance of the inference model is correctly adjusted to the variance of the channel estimation and detection algorithm. 

\section{Conclusion}\label{sec:conclusion}

 We have presented a message passing based algorithm that is able to robustly estimate and track the agent's position in multipath channels based on range and amplitude information of multiple sensors as well as their respective uncertainties in both partial and total \ac{olos} situations.
We analyzed the performance of the algorithm using numerical simulation and showed that the additional information provided by amplitude information as well as by the NLOS object can support the estimation of the agent state and, thus, reduce the number of lost tracks.
%
In partial \ac{olos} situations the performance of the proposed algorithm attained the CRLB (i.e., no lost tracks). 

For this work, we assumed the parameters of the \ac{nlos} \ac{lhf} to be known constants. To overcome this issue, the parameters of the multipath \ac{lhf} \eqref{eq:multipath_lhf} need to be jointly inferred with the agent state. This non-trivial extension requires extended object \ac{pda} \cite{MeyerICASSP2020} and shall be addressed in future work.



%

  \acrodef{awgn}[AWGN]{additive white Gaussian noise}
  \acrodef{bw}[BW]{bandwidth}
  \acrodef{blt}[BLT]{bluetooth}
  \acrodef{cdf}[CDF]{cumulative distribution function}
  \acrodef{crlb}[CRLB]{Cram\'er-Rao lower bound}
  \acrodef{dmc}[DMC]{dense multipath component}
  \acrodef{dut}[DUT]{device under test}
  \acrodef{eirp}[EIRP]{equivalent isotropic radiated power}
  \acrodefplural{esl}[ESLs]{electronic shelf labels} 
  \acrodef{los}[LOS]{line-of-sight}
  \acrodef{mf}[MF]{matched filter}
  \acrodef{ml}[ML]{maximum likelihood}
  \acrodef{mpc}[MPC]{multipath component}
  \acrodef{nlos}[NLOS]{non-\ac{los}}
  \acrodef{pcb}[PCB]{printed circuit board}
  \acrodef{pdf}[PDF]{probability density function}
  \acrodef{reb}[REB]{ranging error bound}
  \acrodef{rss}[RSS]{received signal strength}
  \acrodef{smc}[SMC]{specular multipath component}
  \acrodef{snr}[SNR]{signal-to-noise-ratio}
  \acrodef{sinr}[SINR]{signal-to-interference-plus-noise-ratio}
  \acrodef{tdoa}[TDOA]{time difference of arrival}
  \acrodef{tka}[TKA]{trusted keyless access}
  \acrodef{toa}[TOA]{time-of-arrival}
  \acrodef{aoa}[AOA]{angle-of-arrival}
  \acrodef{uwb}[UWB]{ultra wide band}
  \acrodef{mie}[MIE]{method of interval estimation}
  \acrodef{mc}[MC]{Monte Carlo}
  \acrodef{mse}[MSE]{mean squared error}
  \acrodef{ci}[CI]{confidence interval}
  \acrodef{cl}[CL]{confidence level}
  \acrodef{pdp}[PDP]{power delay profile}
  \acrodef{dps}[DPS]{delay power spectrum}
  \acrodef{dm}[DM]{dense multipath}
  \acrodef{nlike}[NLIKE]{normalized likelihood}
  \acrodef{zzb}[ZZB]{Ziv-Zakai bound}
  \acrodef{ut}[UT]{unscented transform}
  \acrodef{glrt}[GLRT]{generalized likelihood ratio test}
  \acrodef{mse}[MSE]{mean squared error}
  \acrodef{rmse}[RMSE]{root mean squared error}
  \acrodef{nnlike}[NNLIKE]{normalized noise-free likelihood}
  \acrodef{stdv}[STDV]{standard deviation}
  \acrodef{rv}[RV]{random variable}
  \acrodef{bp}[BP]{belief propagation}
  \acrodef{pda}[PDA]{probabilistic data association}
  \acrodef{mp}[MP]{multipath}
  \acrodef{pmf}[PMF]{probability mass function}
  \acrodef{pdaf}[PDAF]{probabilistic data association filter}
  \acrodef{pdaai}[AI-PDAF]{amplitude-information \ac{pdaf}}
  \acrodef{olos}[OLOS]{obstructed \ac{los}}
  \acrodef{spa}[SPA]{sum-product algorithm}
  \acrodef{mmse}[MMSE]{minimum mean-square error}
  \acrodef{lhf}[LHF]{likelihood function}
  \acrodef{fa}[FA]{false alarm}
  \acrodef{ceda}[CEDA]{channel estimation and detection algorithm}   





\bibliographystyle{IEEEtran}
\bibliography{IEEEabrv,references_new}


%
%
%

\end{document}